\documentstyle[12pt]{article}
\begin{document}
\thispagestyle{empty}

\def\ve#1{\mid #1\rangle}
\def\vc#1{\langle #1\mid}

\newcommand{\p}[1]{(\ref{#1})}
\newcommand{\be}{\begin{equation}}
\newcommand{\ee}{\end{equation}}
\newcommand{\sect}[1]{\setcounter{equation}{0}\section{#1}}

\renewcommand{\theequation}{\thesection.\arabic{equation}}

\newcommand{\vs}[1]{\rule[- #1 mm]{0mm}{#1 mm}}
\newcommand{\hs}[1]{\hspace{#1mm}}
\newcommand{\mb}[1]{\hs{5}\mbox{#1}\hs{5}}
\newcommand{\Db}{{\overline D}}
\newcommand{\bea}{\begin{eqnarray}}
\newcommand{\eea}{\end{eqnarray}}
\newcommand{\wt}[1]{\widetilde{#1}}
\newcommand{\und}[1]{\underline{#1}}
\newcommand{\ov}[1]{\overline{#1}}
\newcommand{\sm}[2]{\frac{\mbox{\footnotesize #1}\vs{-2}}
           {\vs{-2}\mbox{\footnotesize #2}}}
\newcommand{\prt}{\partial}
\newcommand{\eps}{\epsilon}

\newcommand{\R}{\mbox{\rule{0.2mm}{2.8mm}\hspace{-1.5mm} R}}
\newcommand{\Z}{Z\hspace{-2mm}Z}

\newcommand{\cd}{{\cal D}}
\newcommand{\cg}{{\cal G}}
\newcommand{\ck}{{\cal K}}
\newcommand{\cw}{{\cal W}}

\newcommand{\vj}{\vec{J}}
\newcommand{\vl}{\vec{\lambda}}
\newcommand{\vz}{\vec{\sigma}}
\newcommand{\vt}{\vec{\tau}}
\newcommand{\vw}{\vec{W}}
\newcommand{\poiss}{\stackrel{\otimes}{,}}

\def\l#1#2{\raisebox{.2ex}{$\displaystyle
  \mathop{#1}^{{\scriptstyle #2}\rightarrow}$}}
\def\r#1#2{\raisebox{.2ex}{$\displaystyle
 \mathop{#1}^{\leftarrow {\scriptstyle #2}}$}}

\renewcommand{\thefootnote}{\fnsymbol{footnote}}
\newpage
\setcounter{page}{0}
\pagestyle{empty}

\vs{8}

\begin{center}

{\LARGE {\bf The exactly integrable systems connected with semisimple 
algebras of the second rank $A_2,B_2,C_2,G_2$}}\\

\vs{8}

{\large A.N. Leznov$^{a,3}$}
{}~\\
\quad \\
{\em {~$~^{(a)}$ IIMAS-UNAM, Apartado Postal 20-726, Meïxico DF
01000, Meï
xico}}\\
{\em {~$~^{(b)}$ Institute for High Energy Physics,}}\\
{\em 142284 Protvino, Moscow Region, Russia}\\
{\em {~$~^{(c)}$ Bogoliubov Laboratory of Theoretical Physics,
JINR,}}\\
{\em 141980 Dubna, Moscow Region, Russia}

\end{center}

\vs{8}

\begin{abstract}

All exactly integrable systems connected with the semisimple algebras
of the second rank with an arbitrary choice of the grading in them
are
presented in explicit form. General solution of such systems are
expressed in 
terms of the matrix elements of two fundamental representations of
corresponding semisimple groups.

\end{abstract}

\vfill

{\em E-Mail:\
leznov@ce.ifisicam.unam.mx }
\newpage
\pagestyle{plain}
\renewcommand{\thefootnote}{\arabic{footnote}}
\setcounter{footnote}{0}

\section{Introduction}

The main goal of the present paper is to demonstrate on the examples
of semisimple algebras of the second order ($ A_2, B_2, C_2, G_2$)
the general construction connecting each semisimple algebra with a
given grading in it with exactly integrable system of equations. The
simplest example of such construction is the two-dimensional Toda
lattice which was considered and integrated for the case of arbitrary
semisimple algebra almost 20 years ago
\cite{LG},\cite{ls0}\footnote{It is necessary to remind that for the
case of $A_n$ series this problem was solved more then 150 years ago
in Darboux's papers}.  Explicit form of exactly integrable systems in
the case of the main grading was found and described in the recent
papers of the author \cite{LM} (so called Abelian case).

In the present paper we follow three different (independent) aims.
Firstly, to introduce novel, unknown up to now exactly integrable
systems, connected with nonabelian gradings (zero order subspace is
non-commutative algebra by itself) (see in this connection the series
of the papers \cite{GC}). Secondly, to get rid of the assumption of
nonabelian Toda theory to use only subspaces with zero and $\pm 1$
graded indexes. And thirdly, and this is not less important, to give
to the reader possibility to use the technique of the group
representation theory (in very restricted volume) and apply it to the
theory of integrable systems. 

In the present paper it is not so important for us to find the
shortest and simplest way to result, but result by itself. So in
concrete examples we try to use only calculations, which the reader
can check by direct computation on the level of rules of
differentiation and combining terms of the same nature.

The paper is organized as follows. In section 2 we present (as a rule
without any proofs) necessary facts and formulae from the
representation theory of the semisimple algebras and groups. Section
3 is devoted to description of the general construction and
mathematical tricks and methods used in the main sections. In the
section 4 concrete examples of semisimple algebras of the second
order are considered in details for all possible gradings. Concluding
remarks and perspectives for further investigation are concentrated
in section 5.

\section{Necessary facts from the representation theory of the
semisimple algebras and groups}

Let ${\cal G}$ be some an arbitrary finite dimensional graded Lie
algebra 
\footnote{We make no difference between algebra and super-algebra
cases, 
recalling only that even (odd) elements of the super-algebras are
always 
multiplied by even (odd) elements of the Grassmann space.}.
This means that $\cal G$ may be represented as a direct sum of
subspaces with
the different grading indexes
\begin{equation}
  {\cal G}=\left(\oplus^{N_-}_{k=1} {\cal G}_{-\frac{k}{2}}\right)
  {\cal G}_0 \left(\oplus^{N_+}_{k=1}{\cal G}_{\frac{k}{2}}\right).
\label{GR}
\end{equation}

The generators with the integer graded indexes are called bosonic,
while with half-integer are the fermionic ones. Positive (negative)
grading corresponds to upper (lower) triangular matrices.

The grading operator $H$ for arbitrary semisimple algebra may be
represented as linear combination of elements of commutative Cartan
subalgebra taking the unity or zero values on the generators of the
simple roots \cite{BUR}
in the form: \begin{equation}
H={\sum}^r_{i=1} (K^{-1}c)_i h_{i}
\label{cartan1}
\end{equation}
Here $K^{-1}$ is the inverse Cartan matrix $K^{-1}K=KK^{-1}=I$
and $c$ is the column consisting of zeros and unities in arbitrary
order. Under the main grading all $c_i=1$ and in this case
$(K^{-1}c)_i=\sum_{j=1} ^r K^{-1}_{i,j}$, $r$ is the rank of the
algebra.

As usual, the generators of the simple roots $X^{\pm}_i$ (raising or
lowering
operators) and Cartan elements $h_i$  satisfy the system of
commutation
relations:
\begin{equation}
[h_i , h_j]=0, \quad [h_i,X^{\pm}_j]=\pm K_{j,i}X^{\pm}_j, \quad
[X^{+}_i,X^{-}_j\}={\delta}_{i,j} h_j, \quad (1 \leq i,j \leq
r),\label{aa6}
\end{equation}
where $K_{ij}$ are the elements of Cartan matrix, and the brackets
$[,\}$ 
denote the graded commutator, $r$- is the rank of the algebra.

The highest vector $\ve{j}$ ($\vc{j} \equiv \ve{j}^{\dagger}$) of
the $j$--th fundamental representation possesses the following
properties:
\begin{eqnarray}
X^{+}_i\ve{j}=0, \quad h_i\ve{j}={\delta}_{i,j}\ve{j}, \quad
\vc{j}\ve{j}=1.
\label{high}
\end{eqnarray}
The representation is exhibited by repeated applications of the
lowering operators $X^{-}_i$ to the $\ve{j}$ and
extracting all linearly-independent vectors with non-zero norm. Its
first
few basis vectors are
\begin{eqnarray}
&& \ve{j}, \quad X^{-}_j\ve{j},  \quad X^{-}_i X^{-}_j\ve{j}\neq
0,\quad
K_{i,j}\neq 0,\quad i\neq j
\label{vectors}
\end{eqnarray}

In the fundamental representations, matrix elements of the
group element $G$ satisfy the following important identity
\footnote{Let us remind the definition of the
superdeterminant, $sdet \left(\begin{array}{cc} A, & B \\ C, & D
\end{array}\right) \equiv det (A-BD^{-1}C ) (det D)^{-1}$.}
\cite{ls0}
\begin{eqnarray}
sdet \left(\begin{array}{cc} \vc{j}X_j^+GX_j^-\ve{j}, &
\vc{j}X_j^+G\ve{j} \\ \vc{j}GX_j^-\ve{j}, & \vc{j} G \ve{j}
\end{array}\right) = {\prod}^r_{i=1}\vc{i} G \ve{i}^{-K_{ji}},
\label{recrel}
\end{eqnarray}
where $K_{ji}$ are the elements of the Cartan matrix. The identity
(\ref
{recrel}) represents the generalization of the famous Jacobi
identity,
connecting determinants of (n-1), n and (n+1) orders of some special
matrices to the case of arbitrary semisimple Lee super-group. As we
will see
in the next section, this identity is so important at the deriving of
the
exactly integrable systems, that one can even say that it is
responsible for
their existence. We conserve for (\ref{recrel}) the name of the first
Jacobi
identity. Besides (\ref{recrel}), there exists no less important
independent identity \cite{l}:
$$
(-1)^P K_{i,j} {\vc{j}X_j^+X_i^+ G \ve{j}\over \vc{j} G \ve{j}}+
K_{j,i} {\vc{i}X_i^+X_j^+ G \ve{i}\over \vc{i} G \ve{i}}+
$$
\begin{equation}
K_{ij}K_{j,i}(-1)^{jP}{\vc{j}X_j^+ G \ve{j}\over \vc{j} G
\ve{j}}{\vc{i}X_i^+
G \ve{i}\over \vc{i} G \ve{i}}=0 ,\quad i\neq j \label{J2}
\end{equation}
which will be called the second Jacobi identity. This identity is
responsible (in the above sense) for the fact of existence of
hierarchy
of integrable systems each one of which is invariant with respect to
transformations of the integrable mapping, connected with each
exactly
integrable system.

Either from (\ref{recrel}), or from (\ref{J2}), it is possible to
construct
many useful recurrent relations which will be used in the further
consideration.

Keeping in the mind the importance of the Jacobi identities
(\ref{recrel}) 
and (\ref{J2}) for further consideration for convenience of the
reader we
present below briefly a proof of (\ref{recrel}).

Let us consider the left hand side of (\ref{recrel}) as a function on
the
group, where $G$ is its arbitrary element. The action on $G$ in the
definite
representation $l$ of the operators of the right (left) regular
representation by definition is as follows:
\begin{equation}
M_{left}(\tilde M_{right}) G= M_l G (\tilde M_l)\label{AM}
\end{equation}
where now $M_l,\tilde M_l$ are generators (the matrices of
corresponding 
dimension) of the shifts on the group in a given $l$ representation.
Now let 
us act on left hand side of (\ref{recrel}) by arbitrary generator of
the 
simple positive root $(X^+_s)_r$. This action is equivalent to 
differentiation and so it is necessary to act consequently on the
first and 
the second columns of the matrix (\ref{recrel}) with summation of the
results.
The action on the second one is always equal to zero as a corollary
of the 
definition of the highest state vector (\ref{high}).
Action on the first column is different from zero only in the case
$s=j$.
But in this case using the same definition of the highest state
vector
we conclude that in the result of differentiation of the first column
it
becomes equal to the second one with the zero final result. So as the
function on the group the left hand side of (\ref{recrel}) is also
proportional to the highest vector ( or the linear combination of
such ones 
vectors) of some other representation. The highest vector of the
irreducible 
representation is uniquely defined by the values, 
which Cartan generators take on it. If Cartan generators take on the
highest 
vector values $V(h_i)=l_i$ than the last one may be uniquely
represented in 
the form:
\begin{equation}
\vc{l} G \ve{l}=C \prod_{i=1}^r(\vc{i} G \ve{i})^{l_i}\label{HVD}
\end{equation}
Calculating the values of Cartan generators on
the left hand side of equation (\ref{recrel}) (both left and right
with
the same result) and using the last comment
about the form of the highest vector we prove (\ref{recrel}) ( the
constant C=1, as one can see by putting $G=1$ and comparing the terms
on the both sides).

The second Jacobi identity can be proved with the help of the
arguments of the
same kind \cite{l}.

The following generalization of the first Jacobi identity will be
very
important in calculations, connected with nonabelian gradings.

Let $\ve{\alpha}$ be the basis vectors of some representation
rigorously
in the order of increasing of the number of the lowering generators
(see 
(\ref{high}) and (\ref{vectors})). We assume also that under the
action of
the generator of arbitrary positive simple root on each basis vector
there 
arises the linear combination of the previous ones.

Then the principal minors of arbitrary order of the matrix 
($G$ is arbitrary element of the group):
$$
G_{\alpha}=\vc{\alpha} G  \ve{\alphaï} 
$$
are annihilated by all generators of positive root from right and
negative 
ones from left.

Indeed this action is equivalent to differentiation and so it is 
necessary to act on each column (line) of the minorïs matrix with 
further summation of the results. But action of the generator of a
positive 
simple root on the state vector with given number of the lowering
operators 
transform it to the state vector with number of lowering operators on
unity 
less, which by condition of the proposition is the linear combination
of 
previous columns (or lines). So in all cases the determinant arises
with 
the linear dependence between it's lines or columns always with the
zero 
result.

The generators of Cartan subalgebra obviously take the definite
values on 
the the minors of such kind and if the corresponding values of them
are 
$l^s_i$, then it is possible to write in correspondence with
(\ref{HVD})
the equality:
\begin{equation}
Min_s= C_s \prod_{i=1}^r \vc{i} G \ve{i}^{l^s_i}\label{GJI}
\end{equation}
and constant $C_s$ may be determined as it was described above.

\section{General construction and technique of computation}

The grading of a semisimple algebra is defined by the values, which
grading
operator $H$ takes on the simple roots of the algebra. As it was
mentioned 
above this values may be only zero and unity ones in an arbitrary
order.
$$
[H, X^{\pm}_i]=\pm X^{\pm}_i,\quad H=\sum_1^r (K^{-1}c)_i h_i,\quad
c_i=1,0
$$

On the level of Dynkin's diagrams the grading can be introduced by
using two 
colors for its dot's: black for simple roots with $c_i=1$ and red
ones 
for the roots with $c_i=0$. With each consequent sequence of the red
(simple) 
roots it is connected the corresponding semisimple algebra (
subalgebra of the
initial one). All such algebras are obviously mutually commutative
and
belong 
to zero graded subspace. To zero graded subspace belong also all
Cartan 
elements of the black roots. We conserve the usual numeration of the
dots of 
Dynkin diagrams and all red algebras will be distinguished by the
index of 
the its first root $m_s$. The rank of $m_s$-th red algebra will be
denoted as 
$R_s$. So $X^{\pm}_{m_s},X^{\pm}_{m_s+1},....X^{\pm}_{m_s+R_s-1}$ are
the 
system of the simple roots of $m_s$ red algebra.

After these preliminary comments we pass to description of the
general 
construction \cite{l1}.

Let two group valued functions $ M^+(y), M^-(x)$ are solutions of
$S$- matrix 
type equations:
\begin{equation}
M^+_y=((\sum_0^{m_2} B^{(+s}(y)) M^+\equiv (B^{(0}+L^+)M^+\quad
M^-_x=M^-(\sum_0^{m_1} A^{(-s}(x))\equiv M^-(A^{(0}+L^-) \label{I}
\end{equation}
where $B^{(+s}(y), A^{(-s}(x)$ take values correspondingly in $\pm s$
graded 
subspaces. $s=0, 1, 2,...m_{1,2}$. In each finite dimensional
representation  
$B^{(+s}(y), A^{(-s}(x)$ are upper (lower) triangular matrices and so 
equations (\ref{I}) are integrated in quadratures.

The key role in what follows plays the composite group valued
function $K$:
\begin{equation}
K=M^+ M^-. \label{II}
\end{equation}
It turns out that matrix elements of $K$ in various fundamental 
representations are connected by closed systems of equivalent
relations, 
which can be interpreted as exactly integrable system with known
general 
solution.

Now we pass to the describing of the necessary calculation methods to
prove 
this proposition.

First of all let us calculate the second mixed derivative $(\ln
\vc{i} K 
\ve{i})_{x,y}$, where index $i$ belongs to the black dot of Dynkin
diagram. 
We have consequently:
\begin{equation}
(\ln \vc{i} K \ve{i})_x= {\vc{i} K (A^0+L^-) \ve{i}\over \vc{i} K
\ve{i}}=
A^0_i(x)+
{\vc{i} K L^- \ve{i}\over \vc{i} K \ve{i}}
\label{MI}
\end{equation}
Indeed, $ K_x=M^+(y)M^-_x(x)=K (A^0+L^-)$, as a corollary of equation
for 
$M^-$. All red components of $A^0$ under the action on the black
highest 
vector state $ \ve{i}$ lead to a zero result in connection with 
(\ref{vectors}). 
The action of Cartan elements of the black roots state vector satisfy
the condition $h_j \ve{i}=\delta_{i,j} \ve{i}$ and so only
coefficient on 
$h_i$ remains in the final result (\ref{MI}).

Further differentiation (\ref{MI}) with respect to $y$, with the 
help of the arguments above, leads to following result:
\begin{equation}
(\ln \vc{i} K \ve{i})_{x,y}=\vc{i} K \ve{i})^{-2} \pmatrix{
\vc{i} K \ve{i}, & \vc{i} K L^- \ve{i} \cr
        \vc{i} L^+ K \ve{i}, &  \vc{i} L^+ K L^- \ve{i}
        \cr}\label{AR}
\end{equation}

Applying (\ref{AM})of the previous section to the left-hand side of
(\ref{AR})
, we obtain finally:
\begin{equation}
(\ln \vc{i} K \ve{i})_{x,y}=L^-_rL^+_l \ln \vc{i} K
\ve{i})^{-1}\label{ARR}
\end{equation}

And thus the problem of the calculation of the mixed second
derivative is 
passed to pure algebraic manipulations on the level of representation
theory 
of semisimple algebras and groups. Further evaluation of (\ref{ARR})
is 
connected with repeated applications of the first (\ref{recrel})
and second (\ref{J2}) Jacobi identities as it will be clear from the
material 
of the next section.

As was mentioned above the red algebras of the zero order graded
subspace in 
general case are not commutative ones and this leads to additional 
computational difficulties. Let us denote by $ \ve{m_i}$
the highest vector of $m_i$-th fundamental representation of the
initial 
algebra. Of course, $\ve{m_i}$ is simultaneously the highest vector
of the 
first fundamental representation of the $m_i$ red algebra. Let
$\vc{\alpha_i}, 
\ve{\beta_i}$ be the basis vectors of the first fundamental
representation 
(this restriction is not essential) of $m_i$-th red algebra and let
us
consider the matrix elements of element $K$ in this basis.
$R_i+1\times R_i+1$ matrix ( $R_i+1$ is the dimension of the first
fundamental 
representation), with the matrix elements  $\vc{\alpha_i} K
\ve{\beta_i}$,
will denoted by a single symbol $u_i$ ( index $i$ takes values from
one to the 
number of the red algebras, which is the function of the chosen
grading).

For derivatives of matrix elements of such constructed matrix we have 
consequently (index $i$ we omit on a moment):
\begin{equation}
\vc{\alpha} u_x \ve{\beta}=\vc{\alpha} K (A^0+L^-) \ve{\beta}=\sum_
{\gamma} \vc{\alpha} K  \ve{\gamma}\vc{\gamma} I A^0 \ve{\beta}
+\vc{\alpha} K L^- \ve{\beta}\label{MCD}
\end{equation}
Or in the equivalent form:
$$
u^{-1} u_x= A^0(x)+u^{-1} \vc{} K L^- \ve{}
$$
Further differentiation with respect to $y$ variable leads to:
$$
\vc{}((u^{-1} u_x)_y\ve{}=u^{-1} \vc{} (B^0+L^+) K L^-
\ve{}-u^{-1}\vc{} 
(B^0+L^+) K \ve{} u^{-1}\vc{} K L^-\ve{}=
$$
\begin{equation}
u^{-1}(\vc{} L^+ K L^- \ve{}-\vc{} L^+ K \ve{} u^{-1} \vc{} K
L^-\ve{})
\label{MC}
\end{equation}
The last expression with the help of standard transformations may be
brought
to the form of the ratio of the two determinants respectively of the 
$R_i+2$ and $R_i+1$ orders:
\begin{equation}
\vc{} u (u^{-1} u_x)_y \ve{}={Det_{N_i+1}\pmatrix{ u & K L^-\ve{} \cr
\vc{} L^+ K  & \vc{} L^+ K  L^- \ve{} \cr}\over Det_{N_i}(u)}
\label{MCC}
\end{equation}

For the discovering of the last expression the generalised Jacobi
identity 
(\ref{GJI}) of the previous section plays the key role and will be
exploited 
many times.

\section{The algebras of the second rank $A_2,B_2C_2,G_2$}

All elements of these algebras may be constructed by consequent
multi-
commutation of generators of four simple roots $X^{\pm}_{1,2}$, with
the 
basic system of commutation relations:
$$ 
[X^+_1,X^-_1]=h_1,\quad
[X^+_1,X^-_2]=[X^+_2,X^-_1]=0,\quad[X^+_2,X^-_2]=h_2
$$
\begin{equation}
[h_1,X^{\pm}_1]= {\pm} 2X^{\pm}_1 \quad [h_2,X^{\pm}_2]= {\pm}
2X^{\pm}_2
\label{RS}
\end{equation}
$$
[h_1,X^{\pm}_2]= {\mp} pX^{\pm}_2 \quad [h_2,X^{\pm}_1]= {\mp}
X^{\pm}_1,
\quad p=1,2,3
$$
In all cases there are possible three nontrivial gradings: $(1,1)$-
the 
principle one (Abelian case), $(1,0)$- the grading of the first
simple root 
and $(0,1)$- of the second simple one. In the case of the principle
grading
corresponding integrable systems for arbitrary semisimple algebras
were found 
and described in \cite{LM}. So each further subsections will be 
devoted to detail considerations of nonabelian gradings
$(1,0),(0,1)$, which
only in the case of $A_2$ algebra are equivalent to each other. 

In the end of this mini-introduction we present the second Jacobi
identity
in application to the algebras of the second rank:
\begin{equation}
{\vc{2} X^+_2 X^+_1 K \ve{2}\over \vc{2} K \ve{2}}+p {\vc{1} X^+_1
X^+_2 K 
\ve{1}\over \vc{1} K \ve{1}}=p {\vc{2} X^+_2  K \ve{2}\over \vc{2} K
\ve{2}}
{\vc{1} X^+_1 K \ve{1}\over \vc{1} K \ve{1}}\label{2JI}
\end{equation}
or in notation, which will be introduced by the way of consideration: 
$$
\bar \alpha_{21}+p\bar \alpha _{12}=p\bar \alpha _1\bar \alpha
_2,\quad
\alpha_{12}+p\alpha _{21}=p\alpha _1 \alpha _2
$$

\subsection{Unitary $A_2$ series}

The root system of this algebra consists of three elements with the
generators $X^{\pm}_1,X^{\pm}_1,X^{\pm}_{12}\equiv \pm
[X^{\pm}_1,X^{\pm}_2]$.
This case corresponds to $p=1$ in (\ref{RS}). For definiteness we
restrict
ourselves by $(1,0)$ grading $[H,X^{\pm}_1]=\mp X^{\pm}_1,
[H,X^{\pm}_2]=0$.

$L^{\pm}$ operators belong to $\pm 1$ graded subspaces and have the
form:
$$
L^+=\bar c_1 X^+_1+\bar c_2 [X^+_2,X^+_1],\quad L^-=c_1 X^-_1+c_2
[X^-_1,
X^-_2]
$$
where $c_{1,2}\equiv c_{1,2}(x),\bar c_{1,2}\equiv \bar c_{1,2}(y)$.

The object of investigation is $2\times 2$ matrix $u$ in the basis of
the
second fundamental representation of $A_2$ algebra \footnote{The
(bra) basis 
vectors of the three dimensional ("quark") second fundamental
representation 
of $A_2$ algebra are the $\vc{2}, \quad \vc{2} X^+_2, \quad \vc{2}
X^+_2 
X^+_1 $.}:
\begin{eqnarray}
u=\left(\begin{array}{cc} \vc{2} K \ve{2}, & \vc{2} K X_2^-\ve{2} \\
\vc{2}X_2^+ K \ve{2}, & \vc{2}X_2^+ K X_2^-\ve{2}
\end{array}\right) \label{recrelI}
\end{eqnarray}
In correspondence with (\ref{MCC}) we have:
\begin{equation}
\vc{} u (u^{-1} u_x)_y \ve{}={Det_3\pmatrix{ u & I K L^- \ve{} \cr
                   \vc{} L^+ K I & \vc{} L^+ K  L^- \ve{} \cr}\over
                   Det_2(u)} 
\label{MCC"}
\end{equation}
The action of operators $L^{\pm}$ on basis vectors $\ve{2}, X^-_2
\ve{2}$
($\vc{2},\vc{2} X^+_2$) is the following:
$$
L^-\ve{2}=c_2 X^-_1 X^-_2 \ve{2}, \quad L^-X^-_2\ve{2}=c_1 X^-_1
X^-_2 \ve{2}
$$
$$
\vc{2}L^+=\bar c_2 \vc{2} X^+_2 X^+_1,\quad  \vc{2} X^+_2 L^+=\bar
c_1 
\vc{2} X^+_2 X^+_1
$$
So in this case the following sequence of basis vectors from
generalized 
Jacobi identity (\ref{GJI}) takes places:
$$
\vc{2}, \quad \vc{2} X^+_2, \quad \vc{2} X^+_2 X^+_1 
$$
The summed values of Cartan generators $h_1,h_2$ on this basis take
zero values and so $Det_3$ from (\ref{MCC"}) equal to unity ( with
correct
account of the constant). This is a really highest vector of scalar,
one- 
dimensional representation of $A_2$ algebra.

Finally (\ref{MCC"}) leads to the system, which matrix function $u$
satisfy: 
\begin{equation}
(u^{-1} u_x)_y =(Det u)^{-1} u^{-1}\pmatrix{ c_2 \bar c_2, & c_1 \bar
c_2  \cr
                   c_2 \bar c_1, & c_2 \bar c_2 \cr} 
\label{A_2}
\end{equation}

In usual notations the system (\ref{A_2}) is nonabelian $A_2 (1,0)$
Toda 
chain. The system (\ref{A_2}) is obviously form-invariant with
respect to  
transformation:
$$
u\to \bar g(y) u \bar g(x)
$$
With the help of this transformation the arbitrary up to now
functions
$c,\bar c$ may be evaluated to a constant values.

\subsection{Orthogonal $B_2$ series equivalent to symplectic one
$C_2$}

This case corresponds to the choice $p=2$ in (\ref{RS}). Both
gradings are 
not equivalent to each other and must be considered separately. First 
fundamental representation for $B_2$ algebra is the second one for
$C_2$ series
and vice versa.

\subsubsection{$(1,0)$ grading}

Generators $L^{\pm}$ may contain components with $\pm 1,\pm2$ graded
indexes and have the form:
$$
L^+=\bar c_1 X^+_1+\bar c_2 [X^+_2,X^+_1]+\bar c^2
[[X^+_2,X^+_1]X^+_1],
$$ 
$$
L^-=c_1 X^-_1+c_2[X^-_1,X^-_2]+c^2 [ X^-_1[X^-_1,X^-_2]]
$$
The object of investigation is two dimensional matrix $u$ in the
basis of
the second fundamental representation of $B_2$ algebra. The main
equation
(\ref{MCC"}) also does not change. The action of $L^{\pm}$ operators
on 
the basis vectors have now the form
\footnote{Five basis vectors of the first fundamental representation
of the 
$B_2$ algebra are the following: $ \ve{2},X^-_2 \ve{2},
X^-_1 X^-_2 \ve{2},X^-_1 X^-_1 X^-_2 \ve{2},X^-_2 X^-_1 X^-_1 X^-_2
\ve{2}$}:
$$
L^-\ve{2}=(c_2+c^2 X^-_1) X^-_1 X^-_2 \ve{2}, \quad L^-X^-_2\ve{2}=
(c_1 +c^2 X^-_2 X^-_1) X^-_1 X^-_2 \ve{2}
$$
$$
\vc{2}L^+=\vc{2} X^+_2 X^+_1(\bar c_2+\bar c^2 X^+_1),\quad  
\vc{2} X^+_2 L^+= \vc{2} X^+_2 X^+_1 (\bar c_1+\bar c^2 X^+_1 X^+_2)
$$

Substituting this expression into (\ref{MCC"}) after some trivial 
evaluations we come to the following relation:
\begin{equation}
u(u^{-1} u_x)_y =(Det u)^{-1} 
\pmatrix{ \bar c_2+\bar c^2 (X^+_1)_l, & 0 \cr
        \bar c_1+\bar c^2 (X^+_1 X^+_2)_l , & 0 \cr} 
\pmatrix{ c_2+c^2 (X^-_1)_r , & c_1 +c^2 (X^-_2 X^-_1)_r \cr
                   0 , &  0 \cr} Det_3 \label{F}
\end{equation}                  
In the last expression $Det_3$ satisfy all conditions of (\ref{GJI}),
with the sequence of bases vectors:
$$
\vc{2}, \quad \vc{2} X^+_2, \quad \vc{2} X^+_2 X^+_1 
$$
In this case the summed value of Cartan element $h_1$ is equal to
2, of $h_2$- to 0. So with the correct value of numerical factor we
obtain $Det_3=2 \vc{1} K \ve{1}^2 $.

The action of the first line operator in (\ref{F}) on $(\vc{1} K
\ve{1})^2$
leads to the line of the form:
\begin{equation}
2 (\vc{1} K \ve{1})^2 (c_2+2 c^2 \alpha_1 , c_1 +2 c^2 \alpha_{21})
\label{L}
\end{equation}
where following abbreviations are used:
$$
\bar \alpha_1={\vc{i} X^+_i K \ve{i}\over \vc{i} K \ve{i}},\quad
\bar \alpha_{12}={\vc{1} X^+_1 X^+_2 K \ve{1}\over \vc{1} K
\ve{1}},\quad
\bar \alpha_{21}={\vc{2} X^+_2 X^+_1 K \ve{2}\over \vc{2} K \ve{2}},
i=1,2
$$
\begin{equation}
\alpha_i={\vc{i} K X^-_i \ve{i}\over \vc{i} K \ve{i}},\quad
\alpha_{21}={\vc{1} K X^-_2 X^-_1 \ve{1}\over \vc{1} K \ve{1}}\quad
\alpha_{12}={\vc{2} K X^-_1 X^-_2 \ve{2}\over \vc{2} K
\ve{2}}\label{NOT}
\end{equation}
Now it is necessary to act with the help of the column operator
(\ref{F})
on the line (\ref{L}). The result of this action on scalar factor may
be 
presented in the form ($Det_2 u=\vc{1} K \ve{1}^2$):
$$
2\pmatrix{ \bar c_2+2 \bar c^2 \bar \alpha_1, & 0 \cr
        \bar c_1+2 \bar c^2 \bar \alpha_{12} , & 0 \cr} 
\pmatrix{ c_2+2 c^2 \alpha_1 , & c_1 +2 c^2 \alpha_{21} \cr
                   0 , &  0 \cr} 
$$
The action of the column operator (\ref{F}) on the line (\ref{L})
leads
to additional matrix:
$$
4c^2 \bar c^2 \pmatrix{ (X^+_1)_l \alpha_1 & (X^+_1)_l \alpha_{21}
\cr
                (X^+_2 X^+_1)_l \alpha_1 & (X^+_2 X^+_1)_l
                \alpha_{21} \cr}
$$
With the help of formulae of AppendixI the last matrix may be
evaluated to 
the form:
$$
4 c^2 \bar c^2 (Det u)^{-1} u .
$$
Gathering all results together, we obtain finally:
\begin{equation}
u(u^{-1} u_x)_y =2 \pmatrix{ p_1 \bar p_1, & p_2 \bar p_1 \cr
                             p_1 \bar p_2, & p_2 \bar p_2 \cr}+ 
4 c^2 \bar c^2 (Det u)^{-1} u \label{B_2}
\end{equation}
where
$$
p_1=c_2+2 c^2 \alpha_1,\quad \bar p_1=\bar c_2+2 \bar c^2 \bar
\alpha_1,\quad
p_2=c_1+2 c^2 \alpha_{21},\quad \bar p_2=\bar c_1+2 \bar c^2 \bar
\alpha_{12}
$$

Now we would like to show that the derivatives $(p_{\alpha})_y$ and
$(\bar p_{\alpha})_x$ are functionally dependent on matrix $u$ and 
themselves, closing in this way the system of equations of
equivalence and 
representing it in the form of closed system of equations for $8$
unknown
functions: 4 matrix elements of $u$ and $4$ components of $2$
two-dimensional 
spinors $p,\bar p$.

Let us follow now the main steps of the necessary calculations.
Using the introduced above technique we have subsequently:
$$
(p_1)_y=2 c^2 (\alpha_1)_y=
{ 2 c^2\over Det(u)} 
Det\pmatrix{ \vc{1} K \ve{1} & \vc{1} K X^-_1\ve{1} \cr
         \vc{1} L^+ K \ve{1} & \vc{1} L^+ K X^-_1 \ve{1} \cr}
$$
The action of $L^+$ on the state vector $\vc{1}$ is the following:
$$
\vc{1}L^+=\vc{1} X^+_1(\bar c_1-\bar c_2 X^+_2-2 \bar c^2 X^+_2
X^+_1)
$$ 
Substituting the last expression in the previous equation and
using the first Jacobi identity for its two first terms (linear in
$\bar c_1,
\bar c_2$) we obtain:
$$
(p_1)_y={ 2 c^2\over Det(u)}(\bar c_1 \vc{2} K \ve{2} - \bar c_2
\vc{2} X^+_2
K \ve{2})- 
$$
\begin{equation}
{ 2 c^2\over Det(u)} Det\pmatrix{ \vc{1} K \ve{1} & \vc{1} K
X^-_1\ve{1} \cr
\vc{1}X^+_1X^+_2 X^+_1  K \ve{1} & \vc{1}X^+_1X^+_2 X^+_1 K X^-_1
\ve{1} \cr}
\label{AE}
\end{equation}
Substituting into the second Jacobi identity (\ref{2JI}) ($p=2$) the
first
one in the form:
$$
\vc{2} K \ve{2}=Det\pmatrix{ \vc{1} K \ve{1} & \vc{1} K X^-_1\ve{1}
\cr
         \vc{1} X^+_1 K \ve{1} & \vc{1} X^+_1 K X^-_1 \ve{1} \cr}
$$
we obtain after some trivial transformations equality for two second
order
determinants:
$$
\pmatrix{ \vc{1} X^+_1 K \ve{1} & \vc{1} X^+_1 K X^-_1\ve{1} \cr
\vc{1}X^+_1X^+_2  K \ve{1} & \vc{1}X^+_1X^+_2 K X^-_1 \ve{1} \cr}=
\pmatrix{ \vc{1} K \ve{1} & \vc{1} K X^-_1\ve{1} \cr
\vc{1}X^+_1X^+_2 X^+_1  K \ve{1} & \vc{1}X^+_1X^+_2 X^+_1 K X^-_1
\ve{1} \cr}
$$
Evaluating the last column of the first determinant with the help of
the first Jacobi identity:
$$
\vc{1} X^+_1 K X^-_1\ve{1}={\vc{2} K \ve{2}+\vc{1} X^+_1 K \ve{1}
\vc{1} K 
X^-_1\ve{1}\over \vc{1} K \ve{1}}
$$
$$
\vc{1} X^+_1 X^+_2 K X^-_1\ve{1}={\vc{2} X^+_2 K \ve{2}+\vc{1} X^+_1
X^+_2 K 
\ve{1} \vc{1} K X^-_1\ve{1}\over \vc{1}  K \ve{1}}
$$
we obtain for it:
$$
\bar \alpha_1 \vc{2} X^+_2 K \ve{2}-\bar \alpha_{12} \vc{2} K \ve{2}
$$
Finally we have:
\begin{equation}
(p_1)_y={ 2 c^2\over Det(u)}(u_{11}\bar p_2 -u_{21}\bar p_1),\quad
(p_2)_y={ 2 c^2\over Det(u)}(u_{12 }\bar p_2 -u_{22}\bar p_1)
\label{UB_2}).
\end{equation}
So (\ref{B_2}), (\ref{UB_2}) and the same system for derivatives of
$(\bar p)
_x$ is the closed system of identities or $B_2(1,0;2,2;c^2,\bar c^2)$
exactly integrable system connected with the $B_2$ semisimple series.
To the best of our knowledge this system was not mentioned in
literature 
before.

{}From the physical point of view the exactly integrable system
(\ref{B_2}),
(\ref{UB_2}) may be considered as a model of interacting charge
${1\over 2}$
particle $(\bar p, p)$ with scalar-vector neutral field $u$.

Putting $ c^2=\bar c^2=0$, we come back to nonabelian Toda lattice
system for 
single matrix valued unknown function $u$.

\subsubsection{$(0,1)$ grading}

Generators $L^{\pm}$ contain only the components with $\pm 1$ graded
indexes and have the form:
$$
L^+=\bar d_1 X^+_2+\bar d_2 [X^+_1,X^+_2]+{1\ over2} \bar d_3
[X^+_1[X^+_1,
X^+_2]],
$$ 
$$
L^-=d_1 X^-_2+d_2 [X^-_2,X^-_1]+{1\over 2} d_3 [ X^-_1[X^-_1,X^-_2]]
$$
With respect to transformation of $1$- red group $A_1$ functions
$d_i(x), (\bar d_i(y))$ are components of three dimensional $A_1$
vectors. 

The object of investigation is two dimensional matrix $u$ in the
basis of
the second fundamental representation of $B_2$ algebra. The main
equation
(\ref{MCC"}) conserves its form. The action of $L^{\pm}$ operators on
the 
basis vectors have now the form 
\footnote{Four basis vectors of the first fundamental of the $C_2$
algebra 
are the following: $ \ve{1},X^-_1 \ve{1},X^-_2 X^-_1 \ve{1},X^-_1
X^-_2 X^-_1 
\ve{1}$.}:
$$
L^-\ve{1}=(d_2-d_3 X^-_1) , \quad L^-X^-_1\ve{1}=
(d_1 -d_2 X^-_1) X^-_2 X^-_1 \ve{1}
$$
$$
\vc{2}L^+=\vc{2} X^+_2 X^+_1(\bar d_2-\bar d_3 X^+_1),\quad  
\vc{2} X^+_2 L^+= \vc{2} X^+_2 X^+_1 (\bar d_1-\bar d_2 X^+_1 X^+_2)
$$

Substituting this expression into (\ref{MCC"}), keeping in mind that
$Det_3$ 
satisfy all conditions of (\ref{GJI}), after some trivial 
evaluations we come to the following relation $(Det_3=\vc{1} K
\ve{1})$:
\begin{equation}
u(u^{-1} u_x)_y =(Det u)^{-1} 
\pmatrix{ \bar d_2-\bar d_3 (X^+_1)_l, & 0 \cr
        \bar d_1-\bar d_2 (X^+_1)_l , & 0 \cr} 
\pmatrix{ d_2-d_3 (X^-_1)_r , & d_1 -d_2 ( X^-_1)_r \cr
                   0 , &  0 \cr} \vc{1} K \ve{1} \label{FF}
\end{equation}                  
Not cumbersome transformation leads the last expression to the
finally form:
\begin{equation}
(u^{-1} u_x)_y =(Det u)^{-1} u^{-1}
\pmatrix{ \bar d_2, & -\bar d_3  \cr
                   \bar d_1, & -\bar d_2 \cr} u 
\pmatrix{ d_2, & d _1  \cr
          -d_3, & -d_2 \cr}  \label{B21}
\end{equation}
(\ref{B21}) is nonabelian Toda chain for $B_2$ algebra with $(0,1)$
grading. To the best of our knowledge it was not considered before.

System (\ref{B21}) is form-invariant with respect to transformation
$u\to \bar g(y) u g(x)$, with the help of which it is possible to
evaluate matrices depending on $x,y$ arguments to constant values. We
omit
here the question about the possible canonical forms of the system 
$B_2(0,1;1,1;\bar d,d)$ (\ref{B21}).

\subsection{The case of $G_2$ algebra}

As it is possible to expect, this case is the most cumbersome. It
corresponds to 
the choice $p=3$ in (\ref{RS}). Firstly, we will consider the case of
$(0,1)$ grading as the most simple one. It is connected with the
7--th
dimensional first fundamental representation of $G_2$ algebra
(group).
The second one connected with $(1,0)$ grading is $14$--th
dimensional.

\subsubsection{ (0,1) grading}

In this case $L^{\pm}$ may contain the components $\pm 1,\pm 2$
graded subspaces and have the form:
$$
L^+=
$$
$$
\bar d_1 X^+_2+ \bar d_2 [X^+_1,X^+_2]+{1\over 2}\bar d_3
[X^+_1[X^+_1,
X^+_2]]+{1\over 6}\bar d_4 [X^+_1[X^+_1[X^+_1,X^+_2]]]+{1\over 3}
\bar d^2 
[X^+_2[X^+_1[X^+_1[X^+_1,X^+_2]]]]
$$
$L^-=(L^+)^T$, where $T$ sign of transposition; with simultaneously
exchanging all coefficients $\bar d\to d$. This operation we will
call as 
"hermitian conjugation". 

Four coefficient functions $d_i, \bar d_i$ on the generators of the
$\pm 1$
graded subspaces in $L^{\pm}$ are united to the ${3\over 2}$
multiplet,
with respect to gauge transformation initiated by group elements
$g_0(x), 
\bar g_0(y)$ belonging to the first red group.

The first fundamental representation of $G_2$ algebra is $7$-th
dimensional
with the basis vectors:
$$
\ve{1}, X^-_1\ve{1}, X^-_2X^-_1\ve{1}, X^-_1X^-_2X^-_1\ve{1},
X^-_1X^-_1
X^-_2X^-_1\ve{1}, 
$$
$$
X^-_2X^-_1X^-_1X^-_2X^-_1\ve{1}, X^-_1X^-_2X^-_1X^-_1X^+_2X^-_1\ve{1}
$$

The action of the operators $L^{\pm}$ on $A_1$ basis of $u$ matrix is
as 
follows:
$$
\vc{1} L^+=\vc{1} X^+_1X^+_2(\bar d_2 -\bar d_3 X^+_1+{1\over 2} \bar
d_4 
X^+_1X^+_1- \bar d^2 X^+_1X^+_1X^+_2)
$$
$$
\vc{1} X^+_1L^+=\vc{1} X^+_1X^+_2(\bar d_1 - \bar d_2 X^+_1+{1\over
2}
\bar d_3 X^+_1X^+_1-\bar d^2 X^+_1X^+_1X^+_2X^+_1)
$$
The action of the operator $L^-$ on $A_1$ basis from the left may be
obtained 
from the last formulae with the help of "hermitian conjugation": 
$$
L^-\ve{1}= (\vc{1} L^+)^T,\quad  L^-X^-_1\ve{1}= (\vc{1}
X^+_1L^+)^T,\quad
\bar d\to d
$$

As in the previous sections the result of calculation of the main
determinant
(\ref{MCC"}) it is possible to present in the form of the product of
column
operator on the line one applied to the highest vector $\vc{1} K
\ve{1}^2$
of the $(2,0)$ representation of $G_2$ algebra ( see Appendix II). 
The line operator form is the following
$$
[d_2-d_3X^-_1+{1\over 2}d_4(X^-_1)^2-{1\over 4}
d^2(2X^-_1X^-_2-3X^-_2X^-_1)
X^-_1,
$$
$$
d_1-d_2X^-_1+{1\over 2}d_3(X^-_1)^2-{1\over
4}d^2X^-_1(2X^-_1X^-_2-3X^-_2
X^-_1)X^-_1]
$$
where $X^-_i\equiv (X^-_i)_r$. Nonusual (compared with the previous
examples)
form of the coefficient on $d^2$ term is the prise for $p=3$ in
(\ref{RS}) 
in the case of $G_2$ algebra.
The column operator is obtained from the line one with the 
help introduced above rules of the "hermitian conjugation".

For rediscovering of last symbolical expression up to the form of
usual 
$2\times 2$ matrix let us introduce two dimensional column vector
$\bar q$ 
the result of the action of the column operator on the highest vector 
$\vc{1} K \ve{1}^2$ divided by itself. The same in the line case will
be 
denoted as $q$. Explicit expressions for the line 
components of $q$ have the form:
$$
q_1=(d_2+{1\over 3}d^2\alpha_{112})-2(d_3+{2\over
3}d^2\alpha_{12})\alpha_1+
(d_4+2\alpha_2d^2)\alpha_1^2,
$$
\begin{equation}
{}\label{Q}
\end{equation}
$$
q_2=(d_1+{1\over 3}d^2\alpha_{1112})-2(d_2+{1\over
3}d^2\alpha_{112})\alpha_1+
(d_3+{2\over 3}d^2\alpha_{12})\alpha_1^2
$$
and with the help of "hermitian conjugation" corresponding
expressions for 
the components for the column $\bar q$.

The result of the action of line operator on the highest vector in
connection with all said above is equal to the numerical line vector
$\vc{1} K \ve{1}^2 (q_1, q_2)$. The action of the column operator on
it may be divided on two steps: the action on the scalar factor
$\vc{1} K \ve{1}^2 $, with the finally matrix $\vc{1} K \ve{1}^2 \bar
q q$ (multiplication by the law the column on the line) and the terms
with partial mutual differentiation of the scalar and the lines
factors. All formulae for concrete calculation of such kind the 
reader can find in Appendix II. It is necessary to pay attention to
the fact, that $X^+_2 q_i=X^-_2 \bar q_i=0$, which one can check
without any difficulties with the help of formulae of the Appendix I.

Gathering all these results we obtain the equation of equivalence for
$u$ function:
\begin{equation}
u(u^{-1}u_x)_y=det^{-1}(u)\sum_{i,j,k,l} u_{ij} u_{kl} \epsilon_{ik} 
\epsilon_{jl}\bar p^{ik} p^{jl}+4 d^2\bar d^2(Det(u))^{-1}
u\label{BE}
\end{equation}
where $u_{ij}$ elements of the matrix $u$, $\epsilon_{ij}$
symmetrical
tensor of the second rank with the components
$\epsilon_{12}=\epsilon_{21}=-1,
\epsilon_{11}=\epsilon_{22}=1$, $\bar p^{ij},p^{ij}$ are two
dimensional 
column and line vectors correspondingly with the components ( the law
of multiplication is the column on the line):
$$
p^{11}=(d_2+{1\over 3} d^2\alpha_{112},d_1+{1\over 3}
d^2\alpha_{1112}),
$$
$$
p^{12}=p^{21}=(d_2+{1\over 3} d^2\alpha_{112},d_3+{2\over 3}
d^2\alpha_{12}),
$$
$$
p^{22}=(d_4+2d^2\alpha_2,d_3+{2\over 3} d^2\alpha_{12})
$$

It remains only to find the derivatives $(\bar p_{ij})_x,(p_{kl})_y
$ and convince ourselves that together with the (\ref{BE}) they
compose the 
closed system of equations of equivalence or exactly integrable 
$G_2(0,1;2,2;\bar d^2,d^2)$ system.

Four components of  $p^{22},p^{11}$ with respect to transformation of
the 
first red algebra compose the ${3\over 2}$ spin-multiplet. So it will
be 
suitable to redenote them by single four-dimensional symbol $p_i$.
And the    
same for "hermitian conjugating" values $\bar p_i$.

Let us follow the calculation of $(\bar p_4)_x=2\bar d^2 (\bar
\alpha_2)_x$. 
The calculation of this the derivative do not different from the 
corresponding computations of section 3 ( see (\ref{AR}) and
(\ref{ARR})).
We have consequently:
\begin{equation}
(\bar \alpha_2)_x=\vc{2} K \ve{2})^{-2} \pmatrix{
\vc{2} K \ve{2}, & \vc{2} K L^- \ve{2} \cr
        \vc{2} X^+_2 K \ve{2}, &  \vc{2} X^+_2 K L^- \ve{2}
        \cr}\label{ARm}
\end{equation}
With the help of the technique used many times before we evaluate the
last 
expression to:
$$
(\bar \alpha_2)_x=L^-_r(X^+_2)_l \ln \vc{2} K \ve{2}=
$$
$$
[d_1-d_2 X^-_1+{1\over 3} d_3 (X^-_1)^2-{1\over 6} d_4 (X^-_1)^3+
d^2([[[X^-_2,X^-_1]X^-_1]X^-_1]-X^-_2(X^-_1)^3)] \theta_2
$$
Using with respect to the last expression formulae of Appendix I we
come to
the system of equations of equivalence for $\bar p$ components of 
${3\over 2}$ multiplet:
$$
(\bar p_4)_x={2\bar d^2\over Det^2 (u)}(p_1 u_{11}^3-3p_2 u^2_{11}
u_{12}+
3p_3 u_{11} u^2_{12}-p_4 u^3_{12})
$$
$$
(\bar p_3)_x={2\bar d^2\over Det^2 (u)}(p_1 u_{11}^2 u_{21}-p_2
(u^2_{11} u_{22}+
2u_{11} u_{21} u_{12})+p_3 (2u_{11} u_{12}u_{21}+u^2 _{12}u_{21})-p_4  
u^2_{12}u_{22})
$$
\begin{equation}
{}\label{BLE}
\end{equation}
$$
(\bar p_2)_x={2\bar d^2\over Det^2 (u)}(p_1 u_{11} u^2_{21}-p_2
(u^2_{21} u_{12}+
2u_{11} u_{21} u_{22})+p_3 (2u_{22} u_{12}u_{21}+u^2 _{22}u_{11})-p_4 
u^2_{22}u_{12})
$$
$$
(\bar p_1)_x={2\bar d^2\over Det^2 (u)}(p_1 u_{21}^3-3p_2 u^2_{21}
u_{22}+
3p_3 u_{21} u^2_{22}-p_4 u^3_{22})
$$

And corresponding system for derivatives $p_y$, which can be obtained
from 
(\ref{BLE}) with the help of "hermitian conjugation".

The symmetry of the constructed exactly integrable $G_2(0,1;2,2;\bar
d^2,d^2)$ 
system (\ref{BE}), (\ref{BLE}) is higher than any possible
expectations.

{}From the physical point of view this system may be considered as
the 
interaction of charge ${3\over 2}$ spin particle ($p,\bar p$) with 
neutral scalar-vector field $u$.

\subsubsection{(1,0) grading}

In this case  $L^{\pm}$ may contain the components $\pm 1,\pm 2,\pm
3$
graded subspaces and have the form:
$$
L^+=\bar c_1 X^+_1+ \bar c_2 [X^+_1,X^+_2]+\bar c^2
[X^+_1[X^+_1,X^+_2]]+
\bar c^3_1 [X^+_1[X^+_1[X^+_1,X^+_2]]]+\bar c^3_2
[X^+_2[X^+_1[X^+_1[X^+_1,
X^+_2]]]]
$$
$L^-=(L^+)^T$, where $T$ sign of transposition ($(X^+_i)^T=X^-_i)$;
with 
simultaneously exchange of all coefficients $\bar c\to c$. This
operation
was called as "hermitian conjugation" in the previous subsection and
we 
conserve here this notation.

As always we begin from the equation of equivalence for two
dimensional
matrix $u$ connected with the second simple root of $G_2$ algebra.
For the decoded of universal equation (\ref{MCC"}) it is necessary
the knowledge of the action of $L^{\pm}$ on the basis. We represent
below only
part of basis vectors of the second fundamental ($14$-th dimensional)
representation of $G_2$ algebra:
$$
\ve{2}, X^-_2\ve{2}, X^-_1X^-_2\ve{2}, X^-_1X^1_2X^-_2\ve{2}, 
X^-_1X^-_1X^-_1X^-_2\ve{2},X^-_2X^-_1X^-_1X^-_2\ve{2}, 
$$
$$
X^-_2X^-_1X^-_1X^-_1X^-_2\ve{2}, X^-_1X^-_2X^-_1X^-_1X^-_2\ve{2},
X^-_1X^-_2X^-_1X^-_1X^-_1X^-_2\ve{2}
$$
The main equations (\ref{I}) are obviously invariant with the respect
to 
the gauge transformation initiated  by $g_0(x),\bar g_0(y)$ elements
of the 
red algebra of the second simple root. With respect to this
transformations
two coefficients of zero $(c^1,\bar c^1)$ and third $(c^3,\bar c^3)$
order graded subspaces are transformed as spinor (anti-) multiplets;
$c^2,\bar  
c^2$ are the scalar ones. With the help of such transformation it is
always 
possible to satisfy the condition $c^3_2=\bar c^3_2=0$ (what is
essential
simplified the calculation) and reconstruct the general case at the
final
step using invariance condition.

The action of the $L^{\pm}$ operators on the basis states of the
second red 
algebra has the form:
$$
\vc{2} L^+=\vc{2} X^+_2X^+_1(-\bar c^1_2 +\bar c^2 X^+_1-\bar c^3_1
X^+_1X^+_1
+\bar c^3_2 (2X^+_1X^+_1X^+_2-3X^+_1X^+_2X^+_1)
$$
$$
\vc{2} X^+_2L^+=\vc{2} X^+_2X^+_1(\bar c^1_1 +\bar c^2 X^+_1X^+_2+
(X^+_1X^+_1X^+_2-3X^+_1X^+_2X^+_1)(\bar c^3_1-\bar c^3_2 X^+_2)
$$
The action of the operator $L^-$ on $A_1$ basis from the left may be
obtained 
from the last formulae with the help of "hermitian conjugation": 
$$
L^-\ve{2}= (\vc{2} L^+)^T,\quad  L^-X^-_1\ve{2}= (\vc{2}
X^+_1L^+)^T,\quad
\bar c\to c
$$

Taking into account arguments of the Appendix II the result of the
calculation 
of determinant of the third order (\ref{MCC"}) may be presented in
the
operator 
column on line form, acting on the highest vector of $(4,0)$
representation 
($3 \vc{1} K \ve{1}^4$) of $G_2$ algebra.

The line ( "hermitian conjugating" column) operators has the form (
in this
expression we put $c^3_2=\bar c^3_2=0$):
$$
(-c^1_2 +c^2 X^-_1-{1\over 2}\bar c^3_1 (X^-_1)^2,\quad
c^1_1 +c^2 X^-_2X^-_1+{1\over 8}c^3_1
(X^-_2X^-_1X^-_1-6X^-_1X^-_2X^-_1)
$$
Further calculations are on the level of accurate application of 
differentiation rules and combination terms of the same nature.
Equation of equivalence for $u$ function have the final form:
\begin{equation}
u(u^{-1}u_x)y=3 det^{{1\over 3}}(u) \bar p^1 p^1+12 det^{-{1\over
3}}(u)
\bar p^2 p^2 u+18 det^{-1}(u) (u\bar c^3) (c^3 u)\label{HV}
\end{equation}
where $p^1$ is the spinor with the components
$p^1=(-c^1_2+4c^2\alpha_1
-6c^3_1\alpha_1^2, c^1_1+4c^2
\alpha_{21}-c^3_1(\alpha_{121}+2\alpha_1\alpha_{21}))$; scalar $p^2=
c^2-3c^3_1\alpha_1$ and corresponding expressions for bar values.

We present the system of equivalence equations without any further
comments:
\begin{equation}
(\bar p^2)_x=-3 det^{-{2\over 3}} \sum_{i,j,k,l}\bar c^3_i
u_{ij}\epsilon_{kl}
p^1_l,\quad (\bar p^1_i)_x=det^{-{2\over 3}}\bar p^2 \sum_{j,k,l}
u_{ij}
\epsilon_{k,l}p^1_l\label{LLL}
\end{equation}
where $\epsilon_{k,l}=-\epsilon_{l,k}$ antisymmetrical tensor of the
second 
rank $\epsilon_{1,2}=-\epsilon_{2,1}=1$.
And, of, course the corresponding system with the derivatives
$p^1_y,p^2_y$.

Physical interpretation of the last system may be connected with
spinor particle interacting with charged scalar $(p^2,\bar p^2)$ and
neutral scalar-vector fields in two dimensions.

\section{Concluding remarks}

In some sense in the present paper the initial idea of Sofus Lie is
realized to introduce continuos groups as power apparatus for solving
of the differential equations.  

On the examples of semisimple groups of the second order we have
decoded this idea and described in explicit form the exactly
integrable systems, general solution of which is possible to obtain
with the help and in the terms of group representation theory. We
have no doubts ( and partially can prove this) that the same
construction is applicable to the case of arbitrary Lie groups and
hope in the nearest future to prove this statement in the whole
volume or to see this proof in the literature.

\noindent{\bf Acknowledgements.}

Author is indebted to the Instituto de Investigaciones en
Matem'aticas Aplicadas y en Sistemas, UNAM for beautiful conditions
for his work. Author friendly thanks N. Atakishiyev for permanent
discussions in the process of working on this paper and big practical
help.

This work was done under partial support of Russian Foundation for
Basic Researche, grant no. 98--01--00330.

\section{Appendix I}

The formulae below are the general ones and have in their foundation
the first Jacobi identity only.

Let us define:         
$$
\theta_j=\prod_{i=1}^r (\vc{i} G \ve{i})^{-K_{ji}}
$$
As a result of differentiation of $\ln \theta_i$, we obtain:
\begin{equation}
(X^-_q)_r \theta_i=-\theta_i  K_{iq} \alpha_q,\quad (X^+_q)_l
\theta_i=
-\theta_i K_{iq} \bar \alpha_q \label{AI1}
\end{equation}
\begin{equation}
(X^-_q)_r \bar \alpha_i=\delta_{q,i} \theta_i,\quad (X^+_q)_l
\alpha_i=
\delta_{q,i} \theta_i \label{AI2}
\end{equation}

In the case of the second order algebras:
\begin{equation}
\theta_1={\vc{2} G \ve{2}\over \vc{1} G \ve{1}^2}
\quad \theta_2={\vc{1} G \ve{1}^p\over \vc{1} G \ve{1}^2}\label{AI3}
\end{equation}

\section{Appendix II}

Let us consider the determinant of the third order the matrix
entries 
of which are coincided with the matrix elements of $G_2$ group
element $K$ 
taken between the bra and the ket three dimensional bases:
\begin{equation}
\vc{1},\vc{1} X^+_1,\vc{1} X^+_2X^+_1X^+_1X^+_2X^+_1,\quad
\ve{1}, X^-_1\ve{1}, X^-_2X^-_1X^-_1X^-_2X^-_1\ve{1}.\label{AII})
\end{equation}

Acting on such determinant by generator $(X^+_2)_r$ and taking  
$(X^-_1X^-_1)_r$ out of its sign we come to the following ket basis:
$$
\ve{1}, X^-_1\ve{1}, X^-_2X^-_1\ve{1}
$$
which in connection with the (\ref{GJI}) tell us that the initial
$Det_3$
(up to the terms annihilated by generators of the positive simple
roots from
right and negative ones from the left) belongs 
to $(2,0)$ ($Vh_1=2,Vh_2=0$) representation of $G_2$ group. For 
initial determinant $Vh_1=1,Vh_2=0$. Each basis vector (see
(\ref{high})) may be obtained with
consequent application of the lowering operators to the highest
vector
($\vc{1} K \ve{1}^2$ in the present case). There are two possibility
to
combination of the lowering operators:
$$
( (AX^-_2X^-_1+BX^-_1X^-_2)X^-_1 )_r
$$ 
and the same expression from the left combination of the raising
generators.
The condition that $Det_3$ is annihilated by generators
$(X^+_1)_r(X^-_1)_l$,
which is a direct corollary of the structure of the bra and ket
bases,
allow to find relation between the constants $3A+2B=0$ and obtain the
expression used in the main text (\ref{Q}) and above.
We obtain the following value for $det_3$ in basis (\ref{AII}):
$$
Det_3={1\over 16}((2X^-_1X^-_2-3X^-_2X^-_1)X^-_1)(X^+_1(2X^+_2X^+_1-
3X^+_1X^+_2)) \vc{1} K \ve{1}^2+\vc{1} K \ve{1}
$$

Below we present necessary formulae for calculation of (\ref{BE}).
We restrict ourselves by $(11)$ component of it.
All "mixed" terms may be gathered in the following form:
$$
-{\bar d^2\over 4}[2(X^+_1X^+_2X^+_1 q_1)-3(X^+_2X^+_1X^+_1
q_1)-8\bar 
\alpha_1(X^+_2X^+_1 q_1)-8\bar \alpha_1 (X^+_1 q_1)-
$$
\begin{equation}
\bar d_3 (X^+_1 q_1)+2\bar d_4 
\bar \alpha_1(X^+_1 q_1)+{\bar d_4\over 2} ((X^+_1)^2
q_1){}\label{AII1}
\end{equation}
Using the definition of vector $q$ (\ref{Q}) and formulae of Appendix
I,
we obtain:
$$
(X^+_1 q_1)=2\theta_1(p^{22}_1\alpha_1-p^{22}_2)\equiv 2\theta_1
P,\quad
X^+_2 P=0,\quad (X^+_2X^+_1 q_1)=2\theta_1 \bar \alpha_2 P,
$$
$$
(X^+_1X^+_1 q_1)=2\theta_1^2 p^{22}_1-4\theta_1 \bar \alpha_1 P,\quad
(X^+_1X^+_2X^+_1 q_1)=2\theta_1(\bar \alpha_{21}-2 \bar \alpha_1 \bar
\alpha_2
)P+2\theta_1^2 \bar \alpha_2 p^{22}_1
$$
$$
(X^+_2X^+_1X^+_1 q_1)=4\theta_1^2 \bar \alpha_2 p^{22}_1+4d^2
\theta_1^2
\theta_2-4\theta_1 \bar \alpha_1 \bar \alpha_2 P-4\theta_1 \bar
\alpha_{12} P.
$$

\end{document}